# Self-organization and chaos in the metabolism of a cell

V.I. Grytsay[a,*], I.V. Musatenko[b]

[a] *Bogolyubov Institute for Theoretical Physics, 14b, Metrolohichna Str., Kyiv 03680, Ukraine,
vgrytsay@bitp.kiev.ua*
[b] *Taras Shevchenko National University of Kyiv, Faculty of Cybernetics, Department of
Computational Mathematics,
64, Volodymyrska Str., Kyiv, Ukraine, ivmusatenko@gmail.com*

*Aim .* To study the dynamics of auto-oscillations arising at the level of enzyme-substrate interaction in a cell and to find the conditions for the self-organization and the formation of chaos in the metabolic process. *Methods*. A mathematical model of the metabolic process of steroids transformation in Arthrobacter globiformis. The mathematical apparatus of nonlinear dynamics. *Results.* The bifurcations resulting in the appearance of strange attractors in the metabolic process are determined. The projections of the phase portraits of attractors are constructed for some chosen modes. The total spectra of Lyapunov's indices are calculated. The structural stability of the attractors obtained is studied. By the general scenario of formation of regular and strange attractors, the structural-functional connections in the metabolic process in the cell are found. Their physical nature is investigated. *Conclusions.* The presented model explains the mechanism of formation of auto-oscillations observed in the Arthrobacter globiformis cells and demonstrates a possibility of the mathematical modeling of metabolic processes for the physical explanation of the self-organization of a cell and its vital activity.

*Keywords:* metabolic process, mathematical model, self-organization, bifurcation, strange attractor, Lyapunov indices.

## INTRODUCTION

One of the basic problems of natural science is the development of a unified theory of the self-organization of the matter. The idea of a cell as a nonlinear open biochemical system is a basis of this trend in the research. It can help to study physical laws of the] formation and life activity of cells.

It is impossible to consider the biochemical evolution realized in the Nature on the whole. Therefore, the following assertion is taken as a basis of our study: "During the biochemical evolution including the self-organization from a primary "broth" on the Earth, only those biochemical processes have been conserved and evolved, which are the predecessors of contemporary metabolic processes running in cells". Thus, it is of primary importance to study the internal dynamics of a vital cell itself. This will allow
clarification of the laws of biochemical evolution on the Earth and the laws of self-organization of a cell as an open nonlinear system.

For this aim, the authors investigated a biochemical process of steroids transformation in the Arthrobacter globiformis cell [1]. Earlier, together with the experimenters, the mathematical model of this process has been constructed. Within this model, various biotechnological modes arising in a bioreactor at the transformation of steroids were described [2-15]. The mathematical model demonstrated a possibility of the appearance of various auto-oscillatory modes in this biotechnological process. In experiments, this phenomenon of oscillatory dynamics was revealed later for the cells that consume other substrata as well [16, 17].

Analogous oscillatory modes were observed in the processes of photosynthesis, glycolysis, variations of the concentration of calcium in cells, oscillations in heart muscle, and other biochemical processes [18-23]. Though the reasons for the appearance of such auto-oscillations can be different, their study allows the step-by-step investigation of the laws of self-organization in biosystems.

In the present work, we state that the reason for the external auto-oscillations observed in the biosystems containing the *Arthrobacter globiformis* cells is the internal auto-oscillations in the cells. Depending on the external conditions, diffusion, etc., the auto-oscillations can change and appear as either auto-oscillations, or as chaos in the biochemical process in the bioreactor. We will consider the auto-oscillations arising at the level of enzyme-substrate interactions and in the respiratory chain. The parameters of such auto-oscillations can hardly be measured in experiments. The auto-oscillations organize themselves in the total metabolic process of cells at autocatalysis.

It is clear that the creation of the most general universal mathematical model of the cell, which would describe a maximally possible number of metabolic processes running in it, is a problem of great importance. However, since it is very difficult to attain the full correspondence of the results obtained from the model to the experimental data, we will restrict ourselves to modeling the most significant experimentally measurable parts of the metabolic process, which were obtained in the experiments.

Though we consider the specific metabolic process, its basic elements, such as the consumption of a substrate, respiratory chain, and positive feedback, reflect the general mechanisms guiding any metabolic process in any cell.

The development of such models for other metabolic processes in a cell will allow the step-by-step study of the process of self-organization on the whole.

Additionally, the metabolic processes are described as various nonlinear interactions, which gives a possibility to investigate a lot of nonlinearities that are inherent in open systems of any nature and eventually form the universal types of self-organization observed in the Nature [24-27].

## MATHEMATICAL MODEL

The general scheme of metabolic processes running in the *Arthrobacter globiformis* cells under the transformation of steroids is presented in Fig. 1. In the article, the part of the metabolic process [3,4] is discussed, where the autocatalysis regime appears.

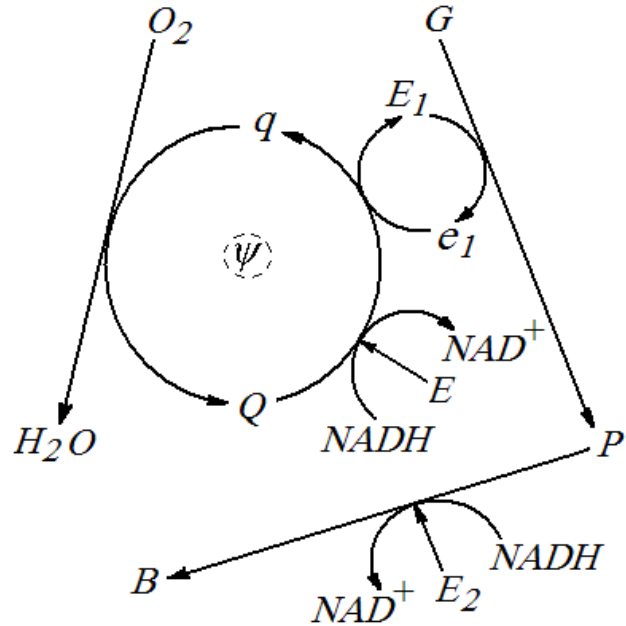

Fig. 1. General scheme of the metabolic process in the *Arthrobacter globiformis* cells.

The mathematical model of the metabolic process in the cell is constructed in accordance with the given general scheme:

$$\frac{dG}{dt} = \frac{G_0}{N_3 + G + \gamma_2 \psi} - l_1 V(E_1) V(G) - \alpha_3 G, \qquad (1)$$

$$\frac{dP}{dt} = l_1 V(E_1) V(G) - l_2 V(E_2) V(N) V(P) - \alpha_4 P, \qquad (2)$$

$$\frac{dB}{dt} = l_2 V(E_2)V(N)V(P) - k_1 V(\psi)V(B) - \alpha_5 B, \tag{3}$$

$$\frac{dE_1}{dt} = E_{10}\frac{G^2}{\beta_1 + G^2}(1 - \frac{P + mN}{N_1 + P + mN}) - l_1 V(E_1)V(G) + l_4 V(e_1)V(Q) - \alpha_1 E_1, \tag{4}$$

$$\frac{de_1}{dt} = -l_4 V(e_1)V(Q) + l_1 V(E_1)V(G) - \alpha_1 e_1, \tag{5}$$

$$\frac{dQ}{dt} = 6lV(2-Q)V(O_2)V^{(1)}(\psi) - l_6 V(e_1)V(Q)_1 - l_7 V(Q)V(N), \tag{6}$$

$$\frac{dO_2}{dt} = \frac{O_{20}}{N_5 + O_2} - lV(2-Q)V(O_2)V^{(1)}(\psi) - \alpha_7 O_2, \tag{7}$$

$$\frac{dE_2}{dt} = E_{20}\frac{P^2}{\beta_2 + P^2}\frac{N}{\beta + N}(1 - \frac{B}{N_2 + B}) - l_{10} V(E_2)V(N)V(P) - \alpha_2 E_2, \tag{8}$$

$$\frac{dN}{dt} = -l_2 V(E_2)V(N)V(P) - l_7 V(Q)V(N) + k_2 V(B)\frac{\psi}{K_{10} + \psi} + \frac{N_0}{N_4 + N} - \alpha_6 N, \tag{9}$$

$$\frac{d\psi}{dt} = l_5 V(E_1)V(G) + l_8 V(N)V(Q) - \alpha\psi. \tag{10}$$

where $V(X) = X/(1+X)$; $V^{(1)}(\psi) = 1/(1+\psi^2)$; $V(X)$ is a function describing the adsorption of the enzyme in the region of its local interaction with the substrate; and $V^{(1)}(\psi)$ is a function characterizing the influence of the kinetic membrane potential on the respiratory chain.

Equations (1)-(9) describe a change in the concentrations of (1)- hydrocortisone $(G)$; (2)- prednisolone $(P)$; (3)- $20\beta$- oxyderivative of prednisolone $(B)$; (4)- oxidized form of 3-ketosteroid-$\Delta^'$-dehydrogenase $(E_1)$; (5)- reduced form of 3-ketosteroid-$\Delta^'$-dehydrogenase $(e_1)$; (6)- oxidized form of respiratory chain $(Q)$; (7)- oxygen $(O_2)$; (8)- $20\beta$-oxysteroid-dehydrogenase $(E_2)$; (9)- $NAD \cdot H$ (reduced form of nicotinamideadeninedinuleotide) $(N)$. Equation (10) describes a change in the kinetic membrane potential $(\psi)$.

The main parameters of the system (1)-(10) were defined from stationary regimes, with appropriate experimental characteristics [3, 4]. Using dimensionless parameters:

$l = l_1 = k_1 = 0.2$; $l_2 = l_{10} = 0.27$; $l_5 = 0.6$; $l_4 = l_6 = 0.5$; $l_7 = 1.2$; $l_8 = 2.4$; $k_2 = 1.5$; $E_{10} = 3$; $\beta_1 = 2$; $N_1 = 0.03$; $m = 2.5$; $\alpha = 0.033$; $a_1 = 0.007$; $\alpha_1 = 0.0068$; $E_{20} = 1.2$; $\beta = 0.01$; $\beta_2 = 1$; $N_2 = 0.03$; $\alpha_2 = 0.02$; $G_0 = 0.019$; $N_3 = 2$; $\gamma_2 = 0.2$; $\alpha_5 = 0.014$; $\alpha_3 = \alpha_4 = \alpha_6 = \alpha_7 = 0.001$; $O_{20} = 0.015$; $N_5 = 0.1$; $N_0 = 0.003$; $N_4 = 1$; $K_{10} = 0.7$.

Solutions of the mathematical model were investigated using the nonlinear differential equations theory [28-30].

The solutions of the system (1)-(10) were obtained, using Runge-Kutta-Merson. The precision of calculations is $10^{-8}$. The time of the entrance on the attractor is $10^6$.

The bifurcation diagrams were built on the plane $G(\alpha)$, using the method of sections. In the phase space the attractor was crossed the plane of section $P = 0.2$. [Бифуркационные диаграммы были построены на плоскости $G(\alpha)$, используя метод сечений. В фазовом пространстве аттрактор был пересечен плоскостью сечения $P = 0.2$] The parameter $G(t)$ was a constant for this intersection of phase trajectory. For the regular attractor, the dots on the plane of section coincide each with other in each of period of the attractor [Для регулярного аттрактора, точки на плоскости сечения совпадают друг с другом в каждом из периодом аттрактора.] For the chaotic attractor, the dots on the plane do not coincide and are chaotic.

The full spectrum of Lyapunov indices was obtained, using Benettine algorithm, with Gram-Schmidt orthogonalization [28]. The types of the attractors were defined, using the first and the second Lyapunov indices. For the regular attractor $\lambda_1 = 0$ and $\lambda_2 < 0$. Then, if $|\lambda_i| < 0.0001$, $\lambda_i \approx 0$.

## RESULTS

Here, we continue to study the dynamics of modes of the mathematical model (1)-(10) at a variation of the dissipation of the kinetic membrane potential $\alpha$. We analyze the various types of auto-oscillatory modes, as well as the scenarios of the appearance of bifurcations under the transition of the dynamical process from one type of attractors to another, with the help of the construction of bifurcation diagrams. As a result of the numerical experiment, we obtained the bifurcation diagrams and calculated the total spectra of Lyapunov indices for the most typical modes (see Table 1).

For the parameter $\alpha = 0.032180$ (Table 1) we can see the transition from the strange attractor $14 \cdot 2^x$ to the strange attractor $13 \cdot 2^x$ (Fig. 2,a). The transition was due to the intermittence. This can be determined by the laminar part of the kinetic curves of attractors. In the system, the transition "chaos-chaos" arises. Up to $\alpha = 0.03222$, only the strange attractor $13 \cdot 2^x$ appeared (Fig. 2,b).

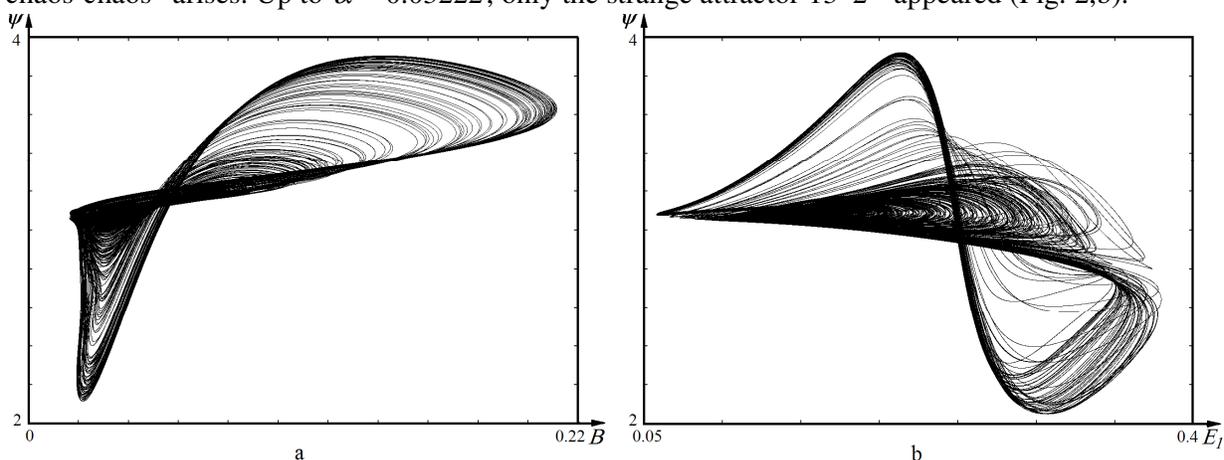

Fig. 2. Projections of the phase portraits:
(a) the strange attractor at the point of the mutual transition $14 \cdot 2^x \leftrightarrow 13 \cdot 2^x$ at $\alpha = 0.032180$;
(b) the strange attractor $13 \cdot 2^x$ at $\alpha = 0.03222$.

Let us consider the following part of the bifurcation diagram: $\alpha \in (0.03227, 0.03234)$ (Fig. 3,a). Looking at the figure from the right to the left we observe that at $\alpha = 0.03234$, there exists a 12-fold limiting cycle on the torus (Fig. 3,c) in the system. A decrease of $\alpha$ causes the appearance of two bifurcations with the doubling of the 12-fold period. Further due to the intermittence, the strange attractor $12 \cdot 2^x$ ($\alpha = 0.03229$, Table 1) arises. It smoothly transits via the intermittence into the strange attractor $13 \cdot 2^x$ ($\alpha = 0.03227575$, Fig. 3b, Table 1).

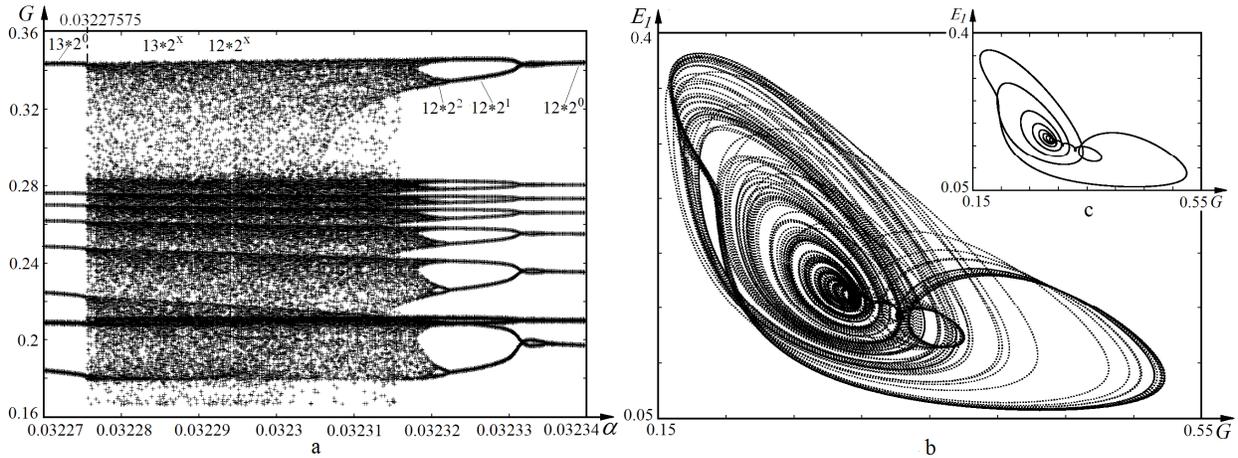

Fig. 3, (a) Bifurcation diagram of the system for $\alpha \in (0.03227, 0.03234)$;

(b) projections of the phase portraits for the strange attractor $13 \cdot 2^x$ ($\alpha = 0.03227575$);

(c) the regular attractor $12 \cdot 2^0$ ($\alpha = 0.03234$).

In Fig. 4, we present the kinetics of the variable $P(t)$ at $\alpha = 0.03229$. We see a 12-fold cycle on the interval $t \in (45, 303)$ and a 13-fold one on the following interval $t \in (303, 579)$. Then the unstable 12-fold cycle is restored again and is replaced even by a 7-fold cycle on the interval $t \in (1579, 1743)$, etc. As a result of the self-organization of the system, we see a gradual transition of the "chaos-chaos" type: from the strange attractor $12 \cdot 2^x$ to the strange attractor $13 \cdot 2^x$.

The transition arose due to the intermittence at the destruction of the laminar part of a trajectory of the 12-fold limiting cycle by the turbulence. In this case, the self-organization of the laminar part into a 13-fold cycle occurs. As $\alpha$ decreases further, we observe the sudden self-organization of the system and the formation of the regular attractor $13 \cdot 2^0$ on the torus ($\alpha = 0.032275$, Table 1) from the strange attractor $13 \cdot 2^x$ ($\alpha = 0.03227575$, Table 1). The "chaos-order" transition arises.

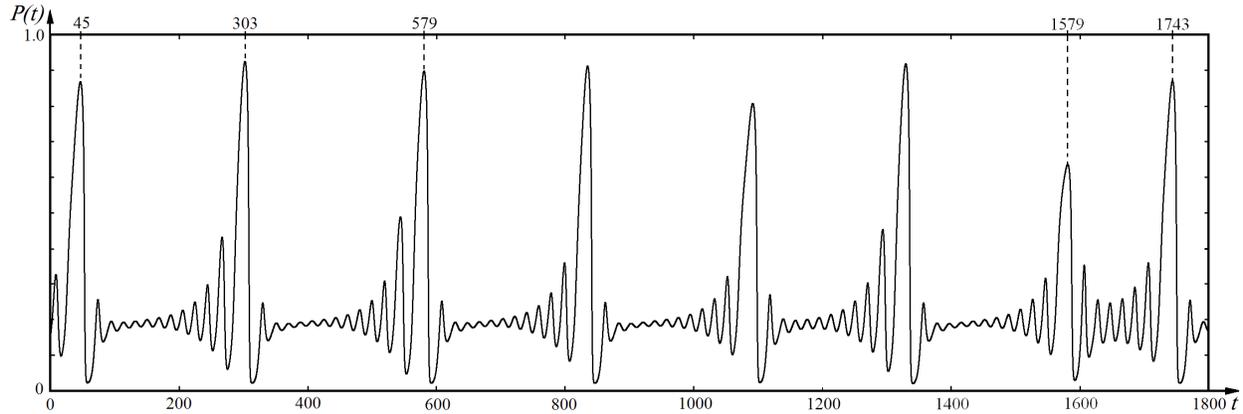

Fig. 4. Kinetic curve for the variable $P(t)$ at $\alpha = 0.03229$.

Below, we consider the parts of the bifurcation diagram with $\alpha \in (0.03238, 0.03245)$ (Fig. 5,a) and $\alpha \in (0.032540, 0.032554)$ (Fig. 5,b).

As $\alpha$ increases from 0.03234 to 0.032386, the regular attractor $12 \cdot 2^0$ transits from a toroidal surface onto a simple 12-fold periodic cycle as a result of the bifurcation accompanied by the disappearance of the torus (Fig. 5,a). On the right end of this part of the bifurcation diagram at $\alpha = 0.03245$, the regular attractor $11 \cdot 2^0$ is present on the torus.

Let us consider the figure from the right to the left. It is very similar to Fig. 3,a. As earlier, a strange attractor is formed via the intermittence from the regular attractor through two bifurcations with the doubling of a period and the intermittence. The "order-chaos" transition $11 \cdot 2^0 \rightarrow 11 \cdot 2^x$ arises. The further decrease of $\alpha$ from 0.0324 down to 0.032387 causes the smooth transition $11 \cdot 2^x \to 12 \cdot 2^x$ of the "chaos-chaos" type via the intermittence. The decrease of $\alpha$ to 0.032386 leads to the self-organization and the formation of the regular attractor $12 \cdot 2^0$.

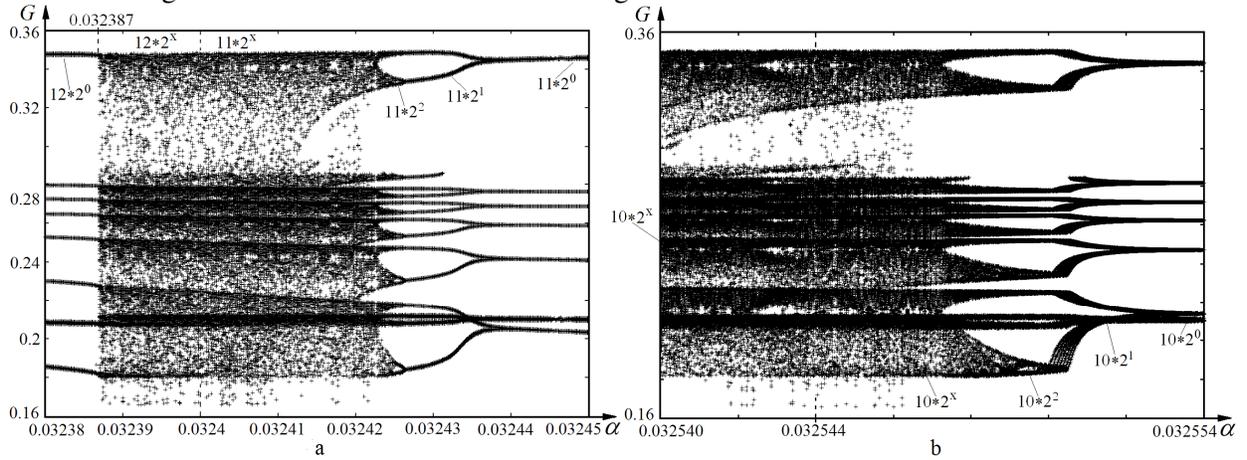

Fig. 5. (a) Bifurcation diagram of the system on the intervals $\alpha \in (0.03238, 0.03245)$;
(b) Bifurcation diagram of the system on the intervals $\alpha \in (0.032540, 0.032554)$.

The further increase of $\alpha$ from 0.03245 to 0.032516 does not affect the configuration of the regular attractor $11 \cdot 2^0$. At $\alpha = 0.032517$, the strange attractor $11 \cdot 2^x$ suddenly arises as a result of the intermittence (Table 1). We observe the transition "order-chaos". As $\alpha$ increases to 0.03254 and more (Fig. 5,b), the laminar part of the trajectory of the attractor $11 \cdot 2^x$ shrinks to $10 \cdot 2^x$, which corresponds to the transition "chaos-chaos".

Considering Fig. 5b from the right to the left, we see the regularity observed earlier: the regular attractor $10 \cdot 2^0$ on the torus ($\alpha = 0.032554$) transits into the strange attractor $10 \cdot 2^x$ ($\alpha = 0.032544$) due to the period doubling bifurcation and the intermittence.

Analyzing Table 1, we can formulate a general scenario of the formation of attractors in the given open nonlinear system. As the dissipation of a kinetic membrane potential decreases from $\alpha = 0.032554$ to $\alpha = 0.0321596$, there occurs the self-organization of the metabolic process in a cell in the following sequence. After the appearance of a definite simple regular attractor, the regular attractor with the same periodicity is formed in the system on the torus after the Neimark bifurcation. Then it is transformed to a strange attractor with the same periodicity as a result of the intermittence. If the dissipation decreases further, the given attractor transits smoothly due to the intermittence into a strange attractor, whose periodicity is greater by 1. The further decrease in the dissipation leads to the self-organization of a regular attractor with the same periodicity from this chaotic attractor. (For example: $11 \cdot 2^0$ ($\alpha = 0.032516$) $\to$ $11 \cdot 2^0(t)$ ($\alpha = 0.03245$) $\to$ $11 \cdot 2^x$ ($\alpha = 0.0324$) $\to$ $12 \cdot 2^x$ ($\alpha = 0.032387$) $\to$ $12 \cdot 2^0$ ($\alpha = 0.032386$)) (Table 1). Then the scenario repeats up to the appearance of a 14-fold strange attractor. Then the behavior of the system becomes somewhat different: there appear additional regular attractors with various periodicities on the torus.

Analyzing the mathematical model, we can determine the structural-functional connections, which makes it possible to explain the appearance of periodic and chaotic oscillations in the metabolic process in a cell. The auto-oscillations arise in the given metabolic process due to the presence of the autocatalytic loop in the system by the accumulation of $NAD \cdot H$ (see (3) and (9)) in the nutrient chain. This yields the desynchronization of the processes: $G \to P$ and $P \to B$, and some periodic cycles are formed. Depending on the kinetic membrane potential $\psi$, the multiplicity of cycles varies from 1-fold to 14-fold. In addition, the variation of $\psi$ affects the activity of the respiratory chain $Q \leftrightarrow q$, and, thus, the second source of bifurcations and auto-oscillations is created. Under the action of the mentioned auto-oscillations, the phase trajectory rotates on the toroidal surface. As a result of the intermittence of the given cycles, a variation of $\alpha$ leads to the formation of quasiperiodic or chaotic modes from a regular attractor.

The further decrease in the dissipation due to the self-organization forms a strange attractor with multiplicity, which is larger than the previous value by 1. The subsequent decrease of $\alpha$ causes the destruction of auto-oscillations arising in the respiratory chain. As a result, only the laminar part of oscillations in the nutrient chain is conserved in the metabolic process, and a new regular attractor arises.

In the subsequent work, we will present the results of the further study of the dynamics of the given metabolic process. We will construct the scenario of the formation of other types of regular and strange attractors, calculate the Lyapunov dimensions of their fractality and KS-entropies, and discuss the "predictability horizons". We will carry out the spectral analysis of the obtained attractors and will study the structure of the chaos of attractors, the hierarchy of its forms, and the influence of the chaos on the stability of the metabolic process and the adaptation and the functioning of a cell.

Table 1. Total spectra of Lyapunov exponents for attractors of the system under study ($\lambda_4$ - $\lambda_9$ are not important for our investigation).

| $\alpha$ | Attractor | $\lambda_1$ | $\lambda_2$ | $\lambda_3$ | $\lambda_4$ - $\lambda_9$ | $\lambda_{10}$ | $\Lambda$ |
|---|---|---|---|---|---|---|---|
| 0.032180 | $(14 \to 13) \cdot 2^x$ | .000928 | -.000023 | -.003588 | --- | -.503950 | -.902093 |
| 0.03222 | $13 \cdot 2^x$ | .000382 | .000016 | -.003923 | --- | -.499280 | -.901762 |
| 0.032275 | $13 \cdot 2^0$ | .000028 | -.000150 | -.003782 | --- | -.500737 | -.904611 |
| 0.03227575 | $13 \cdot 2^x$ | .000575 | -.000013 | -.004226 | --- | -.502724 | -.905510 |
| 0.03229 | $12 \cdot 2^x$ | .000692 | .000061 | -.004093 | --- | -.504396 | -.906552 |
| 0.03234 | $12 \cdot 2^0(t)$ | .000003 | -.000075 | -.004099 | --- | -.502140 | -.907722 |
| 0.032386 | $12 \cdot 2^0$ | .000051 | -.000257 | -.003959 | --- | -.503432 | -.908803 |
| 0.032387 | $12 \cdot 2^x$ | .000601 | -.000069 | -.004380 | --- | -.507443 | -.911089 |
| 0.0324 | $11 \cdot 2^x$ | .000627 | -.000006 | -.004408 | --- | -.507082 | -.911393 |
| 0.03244 | $11 \cdot 2^0(t)$ | -.000021 | -.000030 | -.004336 | --- | -.505113 | -.912083 |
| 0.03245 | $11 \cdot 2^0(t)$ | -.000021 | -.000082 | -.004343 | --- | -.505161 | -.912279 |
| 0.0325 | $11 \cdot 2^0$ | .000019 | -.000669 | -.003864 | --- | -.506524 | -.913908 |
| 0.032516 | $11 \cdot 2^0$ | .000059 | -.000125 | -.004406 | --- | -.507891 | -.914720 |
| 0.032517 | $11 \cdot 2^x$ | .000366 | -.000036 | -.004753 | --- | -.509680 | -.916429 |
| 0.03254 | $10 \cdot 2^x$ | .000274 | -.000016 | -.004579 | --- | -.510670 | -.918002 |
| 0.032544 | $10 \cdot 2^x$ | .000324 | .000015 | -.004567 | --- | -.511698 | -.918904 |
| 0.032554 | $10 \cdot 2^0(t)$ | .000024 | -.000001 | -.004607 | --- | -.509882 | -.918426 |

## CONCLUSIONS

With the help of a mathematical model based on the experimental data, we have studied the process of self-organization in a cell as an open nonlinear system. We have analyzed the auto-oscillatory metabolic processes that are not measurable in experiments and appear on the level of substrate-enzyme interactions. We have determined the parts of bifurcation diagrams with the self-organization and the appearance of chaos. The projections of strange attractors characteristic of the studied metabolic process in a cell are constructed. For the found attractors, the total spectra of Lyapunov indices and the divergences are calculated. The stability of auto-oscillatory processes is studied. The structural-functional connections and the physical nature of instabilities in the metabolic process, which cause the self-organization of the metabolic process, its perturbation, and the adaptation to the change in external conditions, are studied. The auto-oscillations arise due to the autocatalytic loop under the consumption of a substrate by a cell. The quasiperiodic and chaotic types of dynamics appear due to the desynchronization of auto-oscillatory processes with enzyme-substrate interactions and in the respiratory chain. The desynchronization arises under the action of variations of the kinetic membrane potential involved in the metabolic process. The developed model demonstrates the possibility of mathematical modeling of other metabolic processes in cells aimed at the physical explanation of the phenomenon of self-organization observed in a cell in the course of its vital activity.

The work is supported by the project N 0113U001093 of the National Academy of Scienses of Ukraine.

1. *Akhrem A.A., Titov Yu.A.* Steroids and Microorganisms. Moscow: Nauka, 1970. – 526 p.
2. *Gachok V.P., Grytsay V.I.* Kinetic model of macroporous granule with the regulation of biochemical processes. *Dokl. Akad. Nauk SSSR* 1985; **282**: 51-53.
3. *Gachok V.P., Grytsay V.I., Arinbasarova A.Yu., Medentsev A.G., Koshcheyenko K.A., Akimenko V.K.* Kinetic Model of Hydrocortisone 1-en Dehydrogenation by Arthrobacter globiformis. *Biotechn. Bioengin.* 1989; **33**: 661-667.
4. *V.P. Gachok, V.I. Grytsay, A.Yu. Arinbasarova, A.G. Medentsev, K.A. Koshcheyenko, and V.K. Akimenko.* A Kinetic Model for Regulation of Redox Reactions in Steroid Transformation by Arthrobacter globiformis Cells. *Biotechn. Bioengin.* 1989; **33**: 668-680.
5. *Grytsay V.I.* Self-organization in the macroporous structure of the gel with immobilized cells. Kinetic model of bioselective membrane of biosensor. *Dopov. Nats. Akad. Nauk Ukr.* 2000; **2**: 175-179.
6. *Grytsay V.I.* Self-organization in a reaction-diffusion porous media. *Dopov. Nats. Akad. Nauk Ukr.* 2000; **3**: 201-206.
7. *Grytsay V.I.* Ordered structure in a mathematical model biosensor. *Dopov. Nats. Akad. Nauk Ukr.* 2000; **11**: 112–116.
8. *Grytsay V.I.* Self-organization of biochemical process of immobilized cells bioselective of membrane biosensor. *Ukr. J. Phys.* 2001; **1**: 124-127.
9. *Andreev V.V., Grytsay V.I.* Modeling of inactive zones in porous granules katalizatora and biosensor. *Matem. Modelir.* 2005; **17**(2): 57-64.
10. *Andreev V.V., Grytsay V.I.* Influence of heterogeneity of diffusion-reaction process for the formation of structures in the porous medium. *Matem. Modelir.* 2005; **17**(6): 3-12.
11. *Grytsay V.I., Andreev V.V.* The role of diffusion in the active structures formation in porous reaction-diffusion media. *Matem. Modelir.* 2006; **18**(12): 88-94.
12. *Grytsay V.I.* Unsteady Conditions in Porous Reaction-Diffusion. *Medium. Romanian J. Biophys.* 2007; **17**(1): 55-62.
13. *Grytsay V.I.* The uncertainty in the evolution structure of reaction-diffusion medium bioreactor. *Biofiz. Visn.* 2007; **2**: 92 – 97.
14. *Grytsay V.I.* Formation and stability of morphogenetic fields of immobilized cell in bioreactor. *Biofiz. Visn.* 2008; **2**: 25-34.
15. *Grytsay V.I.* Structural Instability of a Biochemical Process. *Ukr. J. Phys.* 2010; **55**(5): 599-606.
16. *Dorofeev A.G., Glagolev M.V., Bondarenko T.F., Panikov N.S.* Unusual growth kinetics of Arthrobacter globiformis and its explanation. *Mikrobiol.* 1992; **61**(1): 33-42.
17. *Skichko A.S., Koltsova E.M.* A mathematical model to describe the fluctuations biomass of bacteria. *Teor. Osnov. Khim. Tekhn.* 2006; **40**(5): 540–550.
18. *Selkov E.E.* Self-Oscillations in Glycolysis. *Europ. J. Biochem.* 1968; **4**: 79-86.
19. *Hess B., Boiteux A.* Oscillatory phenomena in biochemistry. *Annu Rev Biochem.* 1971; **40**: 237–258.
20. *Goldbeter A., Lefer R.* Dissipative structures for an allosteric model. Application to glycolytic oscillations. *Biophys J.* 1972; **12**: 1302-1315.
21. *Godlbeter A., Caplan R.* Oscillatory enzymes. *Annu Rev Biophys Bioeng.* 1976; **5**: 449–476.
22. Chaos in Chemical and Biochemical System / Ed. By R. Field, L. Györgyi. – World Scientific Press, Singapore, 1993.
23. *Kordium V., Irodov D., Maslova O., Ruban T., Sukhorada E., Andrienko V., Shuvalova N., Likhachova L., Shpilova S.* Fundamental biology reached a plateau – development of ideas. *Biopolimers & Cell.* 2011; **27**(6) 480-498.

24. *Turing A.M.* The Chemical Basis of Morphogenesis. *Phil. Trans. R. Soc. Lond.* 1952; **237**(641): 37-72.
25. *Nicolis G., Prigogine I.,* Self-Organization in Nonequilibrium Systems. From Dissipative Structures to Order through Fluctuations. New York: Wiley, 1977.-491 p.
26. *Romanovskii Yu.M., Stepanova N.V., Chernavskii D.S.* Mathematical Biophysics. Moskow: Nauka, 1975.- 305 p.
27. *Akhromeyeva T.S., Kurdyumov S.P., Malinetskii G.G., Samarskii A.A.* Dissipative Structures and Diffusion-Induced Chaos in Nonlinear Media. *Phys. Report.* 1989; **176**: 189-372.
28. *Kuznetsov S.P.* Dynamical Chaos. Moskow: Fiz.-Mat. Nauka., 2001.-296p. (in Russian).
29. *Varfolomeev S.D., Lukovenkov A.V.* Stability in chemical and biological systems: Multistage polyenzymatic reactions. *Russian Journal of Physical Chemistry A.* 2010; **84**(8): 1315-1323.
30. *Anishchenko V.S.* Complex Oscillations in Simple Systems. Moscow: Nauka, 1990. -312 p. (in Russian).

## Самоорганізація і хаос в метаболізмі клітини

### В.Й. Грицай[a,*], І.В. Мусатенко[b]

[a]*Інститут теоретичної фізики ім. М.М. Боголюбова, 14б, вул. Метрологічна, Київ 03680, Україна*

E-mail: vgrytsay@bitp.kiev.ua

[b] *Київський Національний університет ім. Тараса Шевченко, Факультет кібернетики, Кафедра обчислювальної математики, 64, вул. Володимирська, Київ, Україна*

E-mail: ivmusatenko@gmail.com

*Мета. Дослідити динаміку автоколивань, що виникають на рівні фермент-субстратних взаємодій в клітині, знайти умови самоорганізації та утворення хаосу в метаболічному процесі. Методи. Математична модель метаболічного процесу трансформації стероїдів клітиною Arthrobacter globiformis, побудованій по експериментальним даним. Математичний апарат нелінійної динаміки. Результати. Знайдено біфуркації, внаслідок яких в метаболічному процесі виникають дивні атрактори. Для вибраних режимів побудовані проекції фазових портретів атракторів. Розраховані повні спектри показників Ляпунова. Отримані атрактори досліджено на структурну стійкість. Відповідно знайденого загального сценарію формування регулярних та дивних атракторів визначено структурно-функціональні зв'язки в метаболічному процесі клітини. Досліджена їх фізична природа. Висновки. Данна модель пояснює механізм формування автоколивань, спостерігаємих в клітинах Arthrobacter globiformis і демонструє можливість математичного моделювання метаболічних процесів для фізичного пояснення самоорганізації клітини та її життєдіяльності.*

**Ключеві слова:** метаболічний процес, математична модель, самоорганізація, біфуркація, показники Ляпунова.

## Самоорганизация и хаос в метаболизме клетки

### В.И. Грицай[a,*], И.В. Мусатенко[b]

[a]*Институт теоретической физики им. Н.Н. Боголюбова, 14б, ул. Метрологическая, Киев 03680, Украина*

E-mail: vgrytsay@bitp.kiev.ua

[b] *Киевский Национальный университет им. Тараса Шевченко, Факультет кибернетики, Кафедра вычислительной математики, 64, ул. Владимирская, Киев, Украина*

E-mail: ivmusatenko@gmail.com

***Цель.*** *Исследовать динамику автоколебаний, возникающих на уровне фермент-субстратных взаимодействий в клетке, найти условия самоорганизации и образования хаоса в метаболическом процессе.* ***Методы****. Математическая модель метаболического процесса трансформации стероидов клеткой Arthrobacter globiformis, построенная по экспериментальным данным. Математический аппарат нелинейной динамики.* ***Результаты.*** *Найдены бифуркации, вследствие которых в метаболическом процессе возникают странные аттракторы. Для выбранных режимов построены проекции фазовых портретов аттракторов. Рассчитаны полные спектры показателей Ляпунова. Полученные аттракторы исследованы на структурную устойчивость. Согласно найденного общего сценария формирования регулярных и странных аттракторов определены структурно-функциональные связи в метаболическом процессе клетки.. Исследована их физическая природа.* ***Выводы.*** *Данная модель объясняет механизм формирования автоколебаний, наблюдаемых в клетках Arthrobacter globiformis и демонстрирует возможность математического моделирования метаболических процессов для физического объяснения самоорганизации клетки и ее жизнедеятельности.*

**Ключевые слова:** метаболический процесс, математическая модель, самоорганизация, бифуркация, показатели Ляпунова.